\documentclass[conference]{IEEEtran}
\usepackage{amsmath}
\usepackage{amssymb}
\usepackage[dvips]{graphicx}
\usepackage{epsfig}

\newtheorem{lemma:bitenergylow}{Lemma}
\newtheorem{lemma:bitenergyhigh}[lemma:bitenergylow]{Lemma}

\newtheorem{prop:asympcap}{Theorem}
\newtheorem{prop:flashminbitenergy}[prop:asympcap]{Theorem}
\newtheorem{prop:flashbitenergy}[prop:asympcap]{Theorem}
\newtheorem{prop:pasympcap}[prop:asympcap]{Theorem}

\begin{document}


\title{Energy Efficiency Analysis in Amplify-and-Forward and Decode-and-Forward Cooperative Networks}



%
\author{\authorblockN{Qing Chen and Mustafa Cenk Gursoy}
\authorblockA{Department of Electrical Engineering\\
University of Nebraska-Lincoln, Lincoln, NE 68588\\ Email:
chenqing@huskers.unl.edu, gursoy@engr.unl.edu}}


\maketitle

\begin{abstract}\footnote{This work was supported by the National Science Foundation under Grant CCF -- 0546384 (CAREER).}
In this paper, we have studied the energy efficiency of cooperative
networks operating in either the fixed Amplify-and-Forward (AF) or the
selective Decode-and-Forward (DF) mode. We consider the optimization of the $M$-ary
quadrature amplitude modulation (MQAM) constellation size to
minimize the bit energy consumption under given bit error rate (BER)
constraints. In the computation of the energy expenditure, the
circuit, transmission, and retransmission energies are taken into
account. The link reliabilities and retransmission probabilities are
determined through the outage probabilities under the Rayleigh
fading assumption. Several interesting observations with practical
implications are made. For instance, it is seen that while large constellations
are preferred at small transmission distances, constellation size
should be decreased as the distance increases. Moreover, the cooperative gain
is computed to compare direct transmission and cooperative
transmission.
\end{abstract}
\vspace{-.1cm}
\section{Introduction}
In wireless networks, the introduction of relaying provides
higher link reliability when the source-destination link suffers
severe fading. Among different cooperative strategies, the fixed
(AF) and the selective (DF) cooperative techniques are often
employed in cooperative networks. In the fixed AF, the relay doesn't
perform decoding on received signals and always forwards the
amplified received signals to the destination. The selective DF model
differs from the fixed AF in that it will perform Decode-and-Forward
only if its received signal-to-noise ratio (SNR) $\gamma_{s,r}$ from
the source is greater than a threshold $\gamma_{th}$. By comparison, the
fixed AF is easy to implement while the selective DF may be more
complicated on hardware but performs better in terms of bandwidth efficiency. In this
paper, we consider both models and propose maximum ratio combining (MRC) and non-MRC
decoding at the destination and analyze their energy
efficiencies.

Several previous studies in which energy or power efficiency of cooperative transmissions is considered either opt to minimize the total power
when constrained by a given system outage probability, or maximize
the network lifetime with some instantaneous power constraints
\cite{WJ}, \cite{TW}, \cite{CCN}. However, specific
modulations schemes or link layer retransmissions have not been considered and incorporated
in the analysis in such studies.
Since the modulation size affects the transmission rate and
consequently the transmission time, it certainly has a significant
impact on the total energy consumption. When modulation schemes are explicitly considered, power consumption normalized by the transmission rate, i.e., energy per bit, rather than total power consumption should be used in order to provide fair comparison between different modulation schemes. In addition, energy expended in retransmissions, together with circuit energy consumption, should be included in the energy efficiency analysis for more accurate results.

In this paper, motivated by these considerations, we investigate the energy efficiency of cooperative transmissions by jointly considering transmission, circuit, and retransmission energies, and analyzing the bit energy levels achieved by different MQAM schemes. In the system model, we assume a
Rayleigh fading channel, through which the source sends information
packets to the destination and relay by broadcasting. Constrained by a
given BER requirement, we derive the system outage probability.
The relay is assumed to assist the transmission to the destination in either the AF or DF
modes, with or without MRC. Different MQAM sizes are investigated to
find the optimal constellation size in terms of minimum bit energy
consumption.


\vspace{-.2cm}
\section{System Formulation and Channel Assumptions}
We assume a 3-node cooperative network,
\begin{figure}
\begin{center}
\includegraphics[width = 0.35\textwidth]{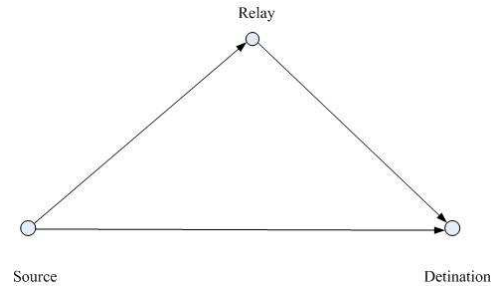}
\caption{One-Relay Cooperative Model} \label{fig:3-node-model}
\end{center}
\end{figure}
where the source has a certain number of bits to transmit to
the destination. While the relay is initially assumed to be located in the middle between the source and
destination, the performance at different locations are analyzed subsequently. The
broadcast from the source can be heard by the relay and destination. When
the destination successfully receives the signals either from the source or
from the relay's cooperative transmission, it sends Acknowledgement back
to the source to indicate a successful transmission. Otherwise, the source
will retransmit until a successful packet delivery is achieved at
the destination. We assume 3 independent Rayleigh fading channels for
the source-destination (S-D) link, the source-relay (S-R) link and
the relay-destination (R-D) link. Moreover, white Gaussian noise
with zero mean and variance $N_{0}$ is assumed to be added to the
received signals at receivers.
\subsection{Cooperative Network Model}
Figure \ref{fig:3-node-model} shows the system model. The
introduction of a relay node definitely adds more energy overhead
and system complexity, but it creates more reliability from the
auxiliary R-D link when S-D channel fails. If we assume the signal
transmitted from the source is $x$ with unit energy and gets broadcasted
through the combined path loss and Rayleigh fading channels, the
received signals at the relay and destination can be represented by
\begin{align}
y_{s,r}=\sqrt{P_{t}d_{s,r}^{-\beta}}h_{s,r}x+n_{s,r}\label{eq:ysr}
\\
y_{s,d}=\sqrt{P_{t}d_{s,d}^{-\beta}}h_{s,d}x+n_{s,d}\label{eq:ysd}
\end{align}
where $P_{t}$ is the transmit power of the source. In the above formulation, $d_{s,r}^{-\beta}$
and $d_{s,d}^{-\beta}$ denote the path loss components as functions of the S-D distance
$d_{s,d}$, the S-R distance $d_{s,r}$ and the path loss exponent
$\beta$. $h_{s,r}$ and $h_{s,d}$ are channel fading coefficients
between S-R and S-D, respectively, modeled as zero-mean circularly
symmetric Gaussian complex random variables. $n_{s,d}$ and $n_{s,r}$
are the additive Gaussian noise at the destination and relay, respectively.

\subsection{Circuit and Transmission Energies}
We adopt the accurate energy consumption formulation from \cite{CM},
by assuming  the source has $L$ bits to transmit directly to
the destination. Such a single transmission consists of 2 distinct
periods: transmission period $T_{on}$ and transient period $T_{tr}$.
Accordingly, the total energy required to send 1 bit is represented
by
\begin{align}
E_{a}=(((1+\alpha)P_{t}+P_{ct}+P_{cr})T_{on}+P_{tr}T_{tr})/L\label{eq:direct}
\end{align}
where $P_{on}=(1+\alpha)P_{t}+P_{ct}+P_{cr}$.

Specifically, $P_{on}$ comprises the transmit power $P_{t}$, the
amplifier power $\alpha P_{t}$, the circuit power $P_{ct}$ at the
source transmitter and $P_{cr}$ at the destination receiver and the
transient power $P_{tr}$.

We consider uncoded square MQAM as our modulation and have
\vspace{-.1cm}
\begin{align}
T_{on}=\frac{LT_{s}}{b}=\frac{L}{bB}\label{eq:Ton}
\end{align}
where $b$ is the constellation size defined as $b=\log_{2}M$ in
MQAM, and $T_s$ is the symbol duration approximated by
$T_{s}\approx1/B$. Finally, the amplifier efficiency for MQAM can be
obtained from $\alpha=\frac{\xi}{\eta}-1$ where
$\xi=3\frac{\sqrt{M}-1}{\sqrt{M}+1}$ in \cite{AD}.

\subsection{BER Constraint for MQAM}
In our model, QoS is constrained by a given BER on each transmission
link and $P_{t}$ is specified. In an uncoded square MQAM, we can
derive the SNR $\gamma_{b}$ according to a given $p_{b}$ in the AWGN
channel, according to (\ref{eq:MQAMprob}) below:
\begin{align}
p_{b}=\frac{1-(1-\frac{2(\sqrt{M}-1)}{\sqrt{M}}Q(\sqrt{\frac{3b\gamma_{b}}{M-1}}))^2}{b}.\label{eq:MQAMprob}
\end{align}
The SNR threshold in our model is therefore derived as
\begin{align}
\gamma_{th}=\frac{P}{N_{0}}=\frac{E_{b}}{N_{0}}\frac{1}{T_{b}}=\gamma_{b}\log_{2}^{M}B.
\end{align}
With $\gamma_{th}$, we can compute the system outage probability in
the following sections.

\section{One-relay AF Cooperative Model with/without MRC}
In fixed AF, network can have 2 different states: in state 1, the source
broadcasts the signal $x$ and the received signals at the relay and
destination can be represented by (\ref{eq:ysr}) and (\ref{eq:ysd}).
At the end of state 1, the destination will first attempt to decode
$y_{s,d}$ to see if it has been received correctly; otherwise the
network will initiate state 2 in which the relay helps to forward
$y_{s,r}$ to the destination by scaling it with $\theta_{r}$
\begin{align}
\theta_{r}=\frac{1}{\sqrt{P_{t}d_{s,r}^{-\beta}|h_{sr}^2|+N_{0}}}
\end{align}
and then transmitting this scaled signal with power $P_{t}$ to the destination. So, the
received signal from the relay at the destination is
\begin{align}
y_{r,d}=\frac{\sqrt{P_{t}d_{r,d}^{-\beta}}}{\sqrt{P_{t}d_{s,r}^{-\beta}|h_{s,r}|^{2}+N_{0}}}h_{r,d}\sqrt{P_{t}d_{s,r}^{-\beta}}h_{s,r}x+n^{'}_{r,d}.\label{eq:yrd}
\end{align}
The noise term in (\ref{eq:yrd}) is
\begin{align}
n^{'}_{r,d}=\frac{\sqrt{P_{t}d_{r,d}^{-\beta}}}{\sqrt{P_{t}d_{s,r}^{-\beta}|h_{s,r}|^{2}+N_{0}}}h_{r,d}n_{s,r}+n_{r,d}\label{eq:rdnoise}
\end{align}
with variance
\begin{align}
N^{'}_{0}=\left(\frac{P_{t}d_{r,d}^{-\beta}|h_{r,d}|^2}{P_{t}d_{s,r}^{-\beta}|h_{s,r}|^{2}+N_{0}}+1
\right)N_{0}.
\end{align}
\subsection{Fixed AF without MRC}
We first study the fixed AF without MRC, where the destination only
utilizes $y_{r,d}$ from the relay for decoding if the S-D link has
failed. After state 2, if the destination still fails to correctly
receive the signal, it then notifies the source to schedule a
retransmission, and this repeats until a successful signal delivery
is achieved at the destination. The successful signal reception
statistically depends on the outage probability of the direct S-D
link and the cooperative S-R-D link. If we assume $|h_{s,r}|^{2}$,
$|h_{s,d}|^{2}$, and $|h_{r,d}|^{2}$ are exponentially distributed with unit
mean, then $\gamma_{s,d}$, which is defined as, 
\begin{align}
\gamma_{s,d}=\frac{P_{t}d_{s,d}^{-\beta}|h_{s,d}|^2}{N_{0}}\label{eq:gamma-sd},
\end{align}
has an exponential distribution with cumulative density function (CDF) evaluated at $\gamma_{th}$ given by
\begin{align}
p(\gamma_{s,d}\leq\gamma_{th})=1-e^{-\frac{N_{0}d_{s,d}^{\beta}\gamma_{th}}{P_{t}}}.\label{eq:gamma-sd-less-gamma-th}
\end{align}
The received SNR $\gamma_{r,d}$ at the destination from the relay is more
complicated and represented by
\begin{align}
\gamma_{r,d}=\frac{(\frac{P_{t}d_{s,r}^{-\beta}}{N_{0}})|h_{s,r}|^2(\frac{P_{t}d_{r,d}^{-\beta}}{N_{0}})|h_{r,d}|^2}{(\frac{P_{t}d_{s,r}^{-\beta}}{N_{0}})|h_{s,r}|^2+(\frac{P_{t}d_{r,d}^{-\beta}}{N_{0}})|h_{r,d}|^2+1}\label{eq:gamma-rd}
\end{align}
and its CDF is formulated as (see e.g., \cite{SA} and \cite{MO})
\begin{align}
p(\gamma_{r,d}\leq\gamma_{th})=1-\sqrt{\xi}e^{-\gamma_{th}(\frac{N_{0}}{P_{t}d_{s,r}^{-\beta}}+\frac{N_{0}}{P_{t}d_{r,d}^{-\beta}})}K_{1}(\sqrt{\xi})
\end{align}
where
$
\xi=\frac{4(\gamma_{th}^{2}+\gamma_{th})N_{0}^{2}}{P_{t}^{2}d_{s,r}^{-\beta}d_{r,d}^{-\beta}}
$ 
and $K_{1}()$ is the first order modified Bessel function of the second kind. The system success probability is therefore derived as
\begin{align}
p_{success1}=1-p_{out}=1-p(\gamma_{s,d}\leq\gamma_{th})p(\gamma_{r,d}\leq\gamma_{th}). \nonumber
\end{align}
With $\gamma_{th}$ derived in Section II, the number of
retransmissions is a geometric random variable and has a mean of
$\frac{1}{P_{success1}}$ \cite{SBAM}. We then formulate the system
average power consumption from
\begin{align}
P_{avg1}&=((1+\alpha)P_{t}+P_{ct}+2P_{cr})p(\gamma_{s,d}\geq\gamma_{th})\nonumber\\
&+(2(1+\alpha)P_{t}+2P_{ct}+3P_{cr})p(\gamma_{s,d}\leq\gamma_{th})\label{eq:power-avg1}
\end{align}
where the power in the first term corresponds to the total
consumption in state 1 where the S-D link is not in outage; the
second term corresponds to both state 1 and state 2 when S-D link is
in outage and the relay participates to transmit the received signal to
the destination. Considering retransmission, the average bit energy
consumption is
\begin{align}
E_{a1}=\frac{P_{avg1}T_{on}+P_{tr}T_{tr}}{Lp_{success1}}.\label{eq:ea1}
\end{align}
To measure how much energy efficiency can be achieved from
cooperative transmission, we define the cooperative gain as
\begin{align}
gain_{1}=\frac{E_{direct}}{E_{a1}}\label{eq:gain1}
\end{align}
where
\begin{align}
E_{direct}=\frac{E_{a}}{p(\gamma_{sd}\geq\gamma_{th})}.\label{eq:power-avg1}
\end{align}

\subsection{Fixed AF with MRC}
The fixed AF with MRC differs from the non-MRC scheme only in state
2: if the S-D link fails, the relay will amplify and forward its
received signal to the destination and we suppose the destination employs a
coherent detector which knows all channel fading coefficients
$h_{s,d}$, $h_{s,r}$ and $h_{r,d}$, such that both $\gamma_{s,d}$
and $\gamma_{r,d}$ will be combined with MRC techniques to output
the maximized equivalent SNR $\gamma$
\begin{align}
\gamma=\gamma_{s,d}+\gamma_{r,d}
\end{align}
where $\gamma_{s,d}$ and $\gamma_{r,d}$ have been derived at
(\ref{eq:gamma-sd}) and (\ref{eq:gamma-rd}).

If the destination still fails to correctly receive the signal in state
2, retransmission will be initiated until a successful signal
delivery at the destination. Therefore, this system has a success probability given by
\begin{align}
p_{success2}&=1-p_{out}=1-p(\gamma_{s,d}\leq\gamma_{th})p(\gamma_{s,d}+\gamma_{r,d}\leq\gamma_{th})
\end{align}
The CDF of $\gamma_{r,d}$ conditioned on $\gamma_{s,d}$ is
\begin{align}
&p(\gamma_{r,d}\leq\gamma_{th}-\gamma_{s,d}|\gamma_{s,d})\nonumber\\
&=1-\sqrt{\xi}e^{-(\gamma_{th}-\gamma_{s,d})(\frac{N_{0}}{P_{t}d_{s,r}^{-\beta}}+\frac{N_{0}}{P_{t}d_{r,d}^{-\beta}})}K_{1}(\sqrt{\xi})
\end{align}
where
\begin{align}
\xi=\frac{4((\gamma_{th}-\gamma_{s,d})^{2}+(\gamma_{th}-\gamma_{s,d}))N_{0}^{2}}{P_{t}^{2}d_{s,r}^{-\beta}d_{r,d}^{-\beta}}.
\end{align}
Then, we average it over $\gamma_{s,d}$ and derive
\begin{align}
&p(\gamma_{r,d}+\gamma_{s,d}\leq\gamma_{th})\nonumber=\\
&\int_{0}^{\gamma_{th}}p(\gamma_{r,d}\leq\gamma_{th}-\gamma_{s,d}|\gamma_{s,d})f(\gamma_{s,d})d\gamma_{s,d}\label{eq:integral}
\end{align}
where
\begin{align}
f(\gamma_{s,d})=\frac{N_{0}}{P_{t}d_{s,d}^{-\beta}}e^{\frac{N_{0}}{P_{t}d_{s,d}^{-\beta}}\gamma_{s,d}}.
\end{align}

We can similarly calculate the average bit energy consumption
$E_{a2}$ and the cooperative $gain_{2}$ in the fixed AF with MRC
similarly as in (\ref{eq:ea1}) and (\ref{eq:gain1}) in the AF
without MRC.

\subsection{Bit Energy Consumption and Cooperative Gain Analysis in AF }
By using the parameters provided in
Table \ref{table:minbitenergy} \cite{CM}, \cite{SSG}, \cite{VE}, we perform numerical computations
on the above two models. The results are shown in the following
figures, with solid-line curves
corresponding to the MRC model and broken-line curves corresponding
to the non-MRC model.
\begin{table}
\caption{Network and Circuit Parameters} \label{table:minbitenergy}
\begin{center}
\begin{tabular}{|c|c|}

\hline $\overline{p}_{b}=10^{-4}$ &$P_{t}$=100mW
\\
\hline $N_{0}=10^{-14}W/Hz$ & $\beta$=3.12
\\
\hline $L$=20000 bits & $freq=2.5*10^{9}Hz$
\\
\hline $P_{tr}$=100mW & $P_{ct}$=98.2mW
\\
 \hline $\eta$=0.35 & $P_{cr}$=112.5mW
\\
\hline $B$=10KHz & $T_{tr}$=5$\mu{s}$
\\
\hline

\end{tabular}
\end{center}
\end{table}

\begin{figure}
\begin{center}
\includegraphics[width = 0.41\textwidth]{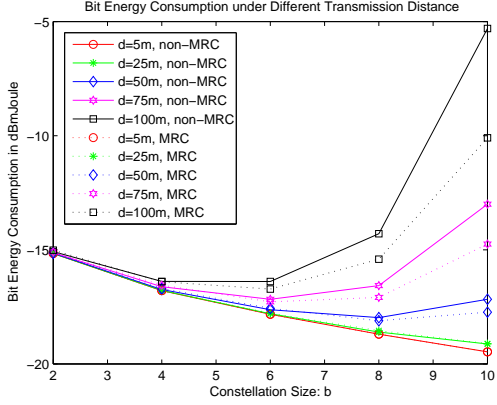}
\caption{Bit Energy Consumption vs Constellation Size in AF}
\label{fig:AF-bit}
\end{center}
\end{figure}

First in Fig. \ref{fig:AF-bit}, we vary the constellation size
$b\in[2,4,6,8,10]$ at a specific S-D distance
$d\in[5m,25m,50m,75m,100m]$ to see how bit energy consumption
changes. We immediately observe the following. At relatively large S-D
distances in either MRC or non-MRC models, the bit energy increases
as $b$ gets either large or very small, and hence, there exists an optimal
$b$ that minimizes the bit energy. We have the following tradeoff.  An MQAM
with large $b$ transmits signals at a faster rate and consequently decreases
$T_{on}$ and hence the energy consumption in a single transmission,
but it also requires a higher $\gamma_{th}$, leading to higher
outage probability and as a result more energy from retransmissions.
A small constellation size requires more energy in a single
transmission, but less retransmissions. Hence, the optimal
constellation size exists to provide a balance between these
effects. We also observe at any S-D distance that both models achieve
the same optimal $b$ and the energy-minimizing constellation size
gets smaller with increasing S-D distance. However, at small
distances such as 5m and 25m, whatever the constellation size is,
the link reliability is always very high. So, large constellation
size is always preferred because it consumes less energy in a single
transmission.

It is immediately seen, as expected, that the MRC model outperforms the non-MRC model
especially at large $b$ and S-D distances. However, at small distances, these
2 models achieve almost the same performance. In such cases, simpler non-MRC techniques can be preferred.

\begin{figure}
\begin{center}
\includegraphics[width = 0.41\textwidth]{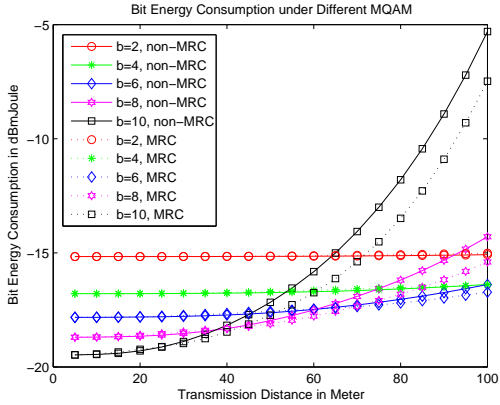}
\caption{Bit Energy Consumption vs S-D Distance in AF}
\label{fig:AF-distance}
\end{center}
\end{figure}

In Fig. \ref{fig:AF-distance}, the bit energy is plotted as a function of S-D
distance at a specific $b\in[2,4,6,8,10]$. We see that while large
constellation sizes are performing well at small distances, small
constellations should be preferred when the distance gets large.
Again, the MRC model outperforms the non-MRC model in terms of lower
bit energy consumption.

\begin{figure}
\begin{center}
\includegraphics[width = 0.41\textwidth]{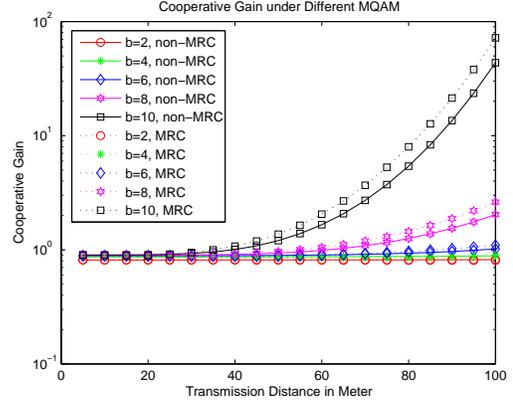}
\caption{Cooperative Gain vs S-D Distance in AF}
\label{fig:AF-distance-gain}
\end{center}
\end{figure}

In Fig. \ref{fig:AF-distance-gain} in which cooperative gain is plotted, the MRC model again achieves
higher gains. We also see when $b\leq6$, the direct transmission
is almost always more energy efficient than the cooperative
transmission. But, the cooperative transmission starts
outperforming the direct transmission at $b=8$ when $d \geq 60m$ and at
$b=10$ when $d\geq 40m$. This is because  at small distances,
cooperation of the relay will increase the energy overhead,
counteracting its energy saving from added reliability and making it
less efficient. On the other hand, at larger $b$ and distances, the energy
saving from less number of retransmissions by improved reliability dominates so
that the system achieves very high cooperative gains.

\begin{figure}
\begin{center}
\includegraphics[width = 0.41\textwidth]{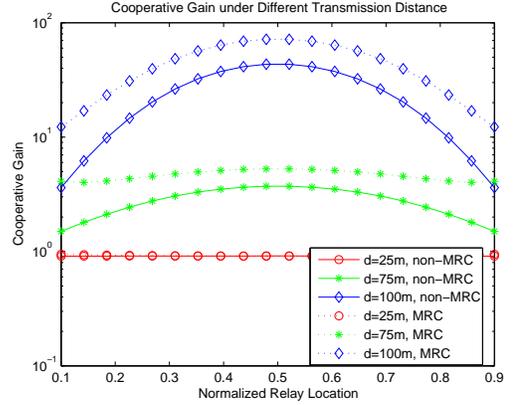}
\caption{Cooperative Gain vs Normalized Relay Location in AF}
\label{fig:AF-location-gain}
\end{center}
\end{figure}

Now, we fix $b=10$ and move the location of the relay either closer to
the source or closer to the destination, indicated by a normalized location
index from 0.1 to 0.9, where 0.5 means the relay is located right in the
middle, in order to see the impact on cooperative gain. In Fig.
\ref{fig:AF-location-gain}, when the S-D distance is 25m, both
non-MRC and MRC achieve the same gain, almost always less than 1. At
higher transmission distances such as 75m and 100m, very high gains
show that the system significantly benefits from the cooperative
transmission and MRC always outperforms non-MRC. Also, we observe
that the maximal cooperative gain is achieved when the relay is located
in exactly the middle of the source and the destination.

\section{One-Relay DF Cooperative Model with/without MRC}
In selective DF, the relay performs decode-and-forward only if
$\gamma_{s,r}\geq\gamma_{th}$ while $\gamma_{s,d}\leq\gamma_{th}$.
So, there are 4 mutually exclusive working states in the DF:
1. S-D link successful; 2. both S-D link and S-R link in outage; 3.
S-D link in outage, but S-R and R-D links successful; 4. S-D link
and R-D in outage, but S-R link successful.

\subsection{Selective DF without MRC}
According to the above 4 different states, the system average power
consumption can be formulated as
\begin{align}
&P_{avg3}=((1+\alpha)P_{t}+P_{ct}+2P_{cr})p(\gamma_{s,d}\geq\gamma_{th})\nonumber\\
&+((1+\alpha)P_{t}+P_{ct}+2P_{cr})
p(\gamma_{s,d}\leq\gamma_{th})p(\gamma_{s,r}\leq\gamma_{th})\nonumber\\
&+(2(1+\alpha)P_{t}+2P_{ct}+3P_{cr})p(\gamma_{s,d}\leq\gamma_{th})
p(\gamma_{s,r}\geq\gamma_{th}).
\end{align}
The system success probability is defined as
\begin{align}
&p_{success3}=p(\gamma_{s,d}\geq\gamma_{th})\nonumber\\
&+p(\gamma_{s,d}<\gamma_{th})p(\gamma_{s,r}\geq\gamma_{th})
p(\gamma_{r,d}\geq\gamma_{th})\label{eq:p-success3}
\end{align}
which can be calculated according to the following CDFs of
exponential distribution
\begin{align}
p(\gamma_{s,d}\leq\gamma_{th})=1-e^{-\frac{N_{0}d_{s,d}^{\beta}\gamma_{th}}{P_{t}}}\label{eq:gammasd}
\end{align}
\begin{align}
p(\gamma_{s,r}\leq\gamma_{th})=1-e^{-\frac{N_{0}d_{s,r}^{\beta}\gamma_{th}}{P_{t}}}\label{eq:gammasr}
\end{align}
\begin{align}
p(\gamma_{r,d}\leq\gamma_{th})=1-e^{-\frac{N_{0}d_{r,d}^{\beta}\gamma_{th}}{P_{t}}}.\label{eq:gammard}
\end{align}
Hence, the bit energy consumption in this model is
\begin{align}
E_{a3}=\frac{P_{avg3}T_{on}+P_{tr}T_{tr}}{Lp_{success3}}\label{eq:ea3}
\end{align}
and the cooperative gain is
\begin{align}
gain_{3}=\frac{E_{direct}}{E_{a3}}.\label{eq:gain3}
\end{align}

\subsection{Selective DF with MRC}
In this model, since we assume the destination employs a coherent
detector, MRC will output a combined
$\gamma=\gamma_{s,d}+\gamma_{r,d}$. The network will have the same 4
working states as the non-MRC DF model with an average power
\begin{align}
P_{avg4}=P_{avg3}.
\end{align}
The system success probability is defined as
\begin{align}
&p_{success4}=p(\gamma_{s,d}\geq\gamma_{th})\nonumber\\
&+p(\gamma_{s,d}<\gamma_{th})p(\gamma_{s,r}\geq\gamma_{th})
p(\gamma_{s,d}+\gamma_{r,d}\geq\gamma_{th}).\label{eq:p-success4}
\end{align}
We know the combined SNR of two independent exponential random
variables $\gamma_{s,d}$ and $\gamma_{r,d}$ has the following
distribution
\begin{align}
f_{z}(z)=&\int_{0}^{z}f_{x}(x)f_{y}(z-x)dx \quad \text{$z>0$}\nonumber\\
=&\frac{\lambda_{a}\lambda_{b}}{\lambda_{b}-\lambda_{a}}[e^{-\lambda_{a}z}-e^{-\lambda_{b}z}]
\end{align}
where $\frac{1}{\lambda_{a}}=\frac{P_{t}}{N_{0}d_{s,d}^{-\beta}}$ is
the mean of $\gamma_{s,d}$ and
$\frac{1}{\lambda_{b}}=\frac{P_{t}}{N_{0}d_{r,d}^{-\beta}}$ is the
mean of $\gamma_{r,d}$.
Therefore, we have
\begin{align}
p(\gamma_{s,d}+\gamma_{r,d}\geq\gamma_{th})=&1-\int_{0}^{\gamma_{th}}f_{z}(z)dz\nonumber\\
=&\frac{\lambda_{b}e^{-\lambda_{a}\gamma_{th}}-\lambda_{a}e^{-\lambda_{b}\gamma_{th}}}{\lambda_{b}-\lambda_{a}}.\label{eq:combined-snr-DF}
\end{align}
We can substitute (\ref{eq:gammasd}), (\ref{eq:gammasr}) and
(\ref{eq:combined-snr-DF}) into (\ref{eq:p-success4}) and similarly
calculate the average bit energy consumption $E_{a4}$ and the
cooperative $gain_{4}$ in the DF with MRC as in
(\ref{eq:ea3}), (\ref{eq:gain3}) in the case of DF without MRC.

\subsection{Bit Energy Consumption and Cooperative Gain Analysis in DF }

\begin{figure}
\begin{center}
\includegraphics[width = 0.41\textwidth]{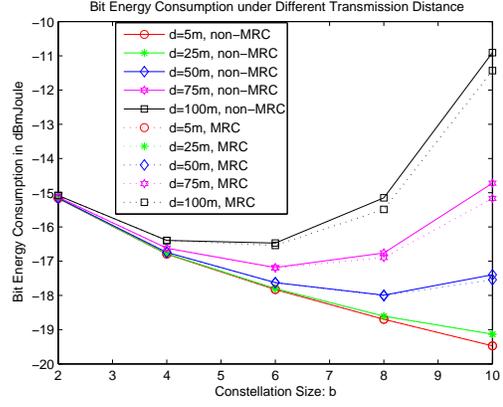}
\caption{Bit Energy Consumption vs Constellation Size in DF}
\label{fig:DF-bit}
\end{center}
\end{figure}

Fig. \ref{fig:DF-bit} shows the bit energy consumption with respect
to $b\in[2,4,6,8,10]$  at a specific distance $d\in[5m,25m,50m,75m,100m]$.
Results similar to those already identified in the fixed
AF model are observed. 
At small S-D distances, the highest constellation should be
preferred. The optimal constellation size gets smaller as the
distance increases. It is noticed that MRC in the DF model
provides only very limited improvement in energy efficiency compared with
non-MRC. Only at points with large $b$ and S-D distances, MRC curves
show better energy efficiency over non-MRC. However, compared with
Fig. \ref{fig:AF-bit} in the fixed AF, the bit energy consumption on
these points has been significantly decreased in the selective DF
model.

Very close performance of MRC and non-MRC in the DF models is also
illustrated in the following figures. In Fig. \ref{fig:DF-distance},
at a given $b$, the bit energy consumption is increasing as the S-D
distance increases, due to the increasing system outage. But, the bit
energy consumption when $b=10$ at large distance has been
significantly decreased in the selective DF models compared with
Fig. \ref{fig:AF-distance} in the fixed AF.
\begin{figure}
\begin{center}
\includegraphics[width = 0.41\textwidth]{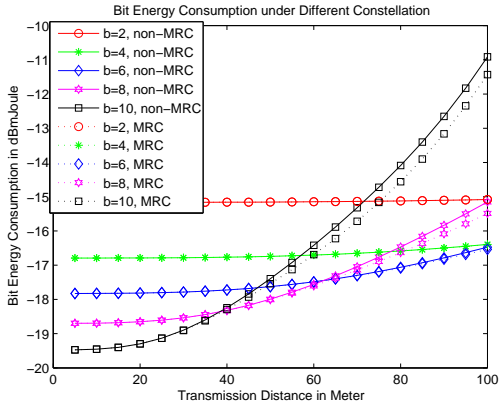}
\caption{Bit Energy Consumption vs S-D Distance in DF}
\label{fig:DF-distance}
\end{center}
\end{figure}

\begin{figure}
\begin{center}
\includegraphics[width = 0.41\textwidth]{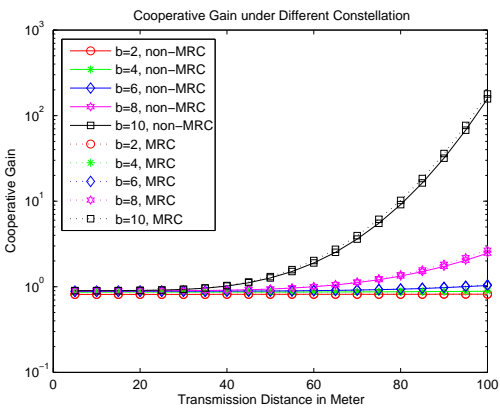}
\caption{Cooperative Gain vs S-D Distance in DF}
\label{fig:DF-distance-gain}
\end{center}
\end{figure}

Fig. \ref{fig:DF-distance-gain} shows that both MRC and non-MRC in the DF mode
almost achieve equal cooperative gain. At $b\in[2,4,6]$, the curves
show almost flat gain less than 1, which advocates that the direct
transmission is more energy efficient. At higher constellation sizes such
as $b=8$ or $b=10$, the system can achieve very high cooperative
gains as the distance increases. Compared with
Fig. \ref{fig:AF-distance-gain}, the DF models are more energy
efficient than the AF models, especially at large $b$ and S-D
distance.

\begin{figure}
\begin{center}
\includegraphics[width = 0.41\textwidth]{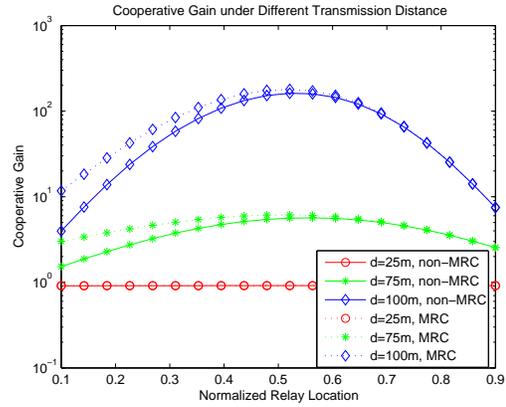}
\caption{Cooperative Gain vs Normalized Relay Location in DF}
\label{fig:DF-location-gain}
\end{center}
\end{figure}

In Fig. \ref{fig:DF-location-gain},  we have fixed $b=10$. When the S-D
distance is 25m in both MRC and non-MRC, changing the relay location
doesn't affect the gain and direct transmission is more energy
efficient because of $gain<1$. At higher S-D distances such as 75m
and 100m, we observe that the system benefits from the cooperative
transmission and achieves relatively high cooperative gains.  MRC
provides higher gain when the relay is closer to the source. As the relay moves
closer to the destination, both MRC and non-MRC models tend to achieve
the same performance. The maximal cooperative gain is again achieved
when the relay is located right in the middle.

\section{Conclusion}
We have considered fixed AF and selective DF models in a 3-node cooperative network. In each model, accurate
energy expenditures which consider the circuit, transmission, and
retransmission energies are formulated. The system reliability is derived
from the link outage probabilities under a combined path-loss and
Rayleigh fading channel. Both MRC and non-MRC decoding are performed
at the destination in AF and DF, and the optimal constellation sizes are
identified. Several interesting results are observed: 1) in fixed AF,
MRC outperforms non-MRC in terms of achieving less bit energy and higher
cooperative gain; 2) in selective DF, MRC doesn't show much improvement
on energy efficiency over non-MRC; 3) at small constellation sizes,
direct transmission is more energy efficient in both models, while
at large constellation sizes, the system can achieve significant
cooperative gains as distance increases; 4) the optimal relay location
is the middle between the source and the destination; 5) the selective DF is
more energy efficient than the fixed AF, especially at large
constellation size and transmission distance.


\begin{thebibliography}{99}


\bibitem{WJ} W.-J. Huang, Y.-W. Perter Hong, C.-C. Jay Kuo, ``Lifetime
Maximization for Amplify-and-Forward Cooperative Networks'', IEEE
Trans. Wireless Commun., vol.7 no.5 pp.1800-1805, May 2008

\bibitem{TW} T. H. Himsoon, W. P. Siriwongpairat, Z. Han, K. J. Ray
Liu,  ``Lifetime Maximization via Cooperative Nodes and Relay
Deployment in Wireless Networks'', IEEE \emph{J.Select. Areas
Commun,} Vol.25  no. 2 pp.302-317, Feb. 2007

\bibitem{CCN} K. J. Ray Liu, A. K . Sadek, W. Su, A. Kwasinski,
``Cooperative Communications and Networking'', Cambridge University
Press, 2009


\bibitem{CM} Q. Chen, M. C. Gursoy, ``Energy Efficient Modulation
Design for Reliable Communication in Wireless Networks'', Conference
on Information Sciences and Systems (CISS), Johns Hopkins University, Mar. 2009

\bibitem{AD}A. J. Goldsmith
Wireless Communications, Cambridge, UK: Cambridge University
Press, 2005.

\bibitem{SA} S. A. Mousavifar, T. Khattab, C. Leung, ``A Predictive
Strategy for Lifetime Maximization in Selective Relay Networks'',
IEEE Sarnoff Symposium, Mar. 2009

\bibitem{MO} M. O. Hasna, M. S. Alouini, ``End-to-End Performance of
Transmission Systems with Relay over Rayleigh-Fading Channel'', IEEE
Trans. Wireless Commun, Vol. 2,No. 6, Nov 2003

\bibitem{SBAM}S. Banerjee and A. Misra, ``Minimum Energy Paths for
Reliable Communication in Multi-hop Wireless networks,'' MOBIHOC,
pp.146-156, June. 2002.

\bibitem{SSG}S. S. Ghassemzadeh, L. J. Greenstein, A. Kavcic, T.
Sveinsson, and V. Tarokh ``UWB Indoor Path Loss Model for
Residential and Commercial Buildings,'' Proc. of VTC 2003, pp. 3115-
3119 Vol.5, Oct 2003.

\bibitem{VE}V. Erceg, L. J. Greenstein, S. Y. Tjandra, S. R. Parkoff, A. Gupta, B. Kulic,
A. A. Julius, and R. Bianchi, ``An Emperically Based Path Loss Model
for Wireless Channels in Suburban Environments,'' IEEE
\emph{J.Select. Areas Commun,} vol. 17, no. 7, pp.1205-1211, Jul.
1999.


\end{thebibliography}
\end{document}